\RequirePackage{fix-cm}
\documentclass{svjour3}                     
\smartqed  
\usepackage{graphicx}
\usepackage{color}
 \journalname{TCA}
\begin{document}

\title{Roaming dynamics in Ketene isomerization 
}


\author{Fr\'ed\'eric A. L. Maugui\`{e}re \and Peter Collins \and Gregory S. Ezra \and Stavros C. Farantos \and Stephen Wiggins}


\institute{F. A. L. Maugui\`ere \at
              School of Mathematics, University of Bristol, Bristol BS8 1TW, United Kingdom \\
              \email{frederic.mauguiere@bristol.ac.uk}           
           \and
           P. Collins \at
              School of Mathematics, University of Bristol, Bristol BS8 1TW, United Kingdom \\
              \email{Peter.Collins@bristol.ac.uk}
          \and
          G. S. Ezra \at
          Department of Chemistry and Chemical Biology, Baker Laboratory, Cornell University, Ithaca, NY 14853 USA \\
          \email{gse1@cornell.edu}
          \and
          S. C. Farantos \at
          Institute of Electronic Structure and Laser, Foundation for Research and Technology - Hellas, 
	and Department of Chemistry, University of Crete, Iraklion 711 10, Crete, Greece \\
	\email{farantos@iesl.forth.gr}
	\and
	S. Wiggins \at
	School of Mathematics, University of Bristol, Bristol BS8 1TW, United Kingdom \\
	\email{stephen.wiggins@mac.com}
}

\date{Received: date / Accepted: date}

\maketitle

\begin{abstract}
A reduced two dimensional model is used to study Ketene isomerization reaction. In light of recent results 
by Ulusoy \textit{et al.} [J.\ Phys.\ Chem.\ A {\bf 117}, 7553 (2013)],
the present work focuses on the generalization of the roaming mechanism to 
the Ketene isomerization reaction by applying
our phase space approach previously used to elucidate the roaming phenomenon in ion-molecule reactions.
Roaming is again found be associated with 
the trapping of trajectories in a phase space region between two dividing surfaces; trajectories are classified as reactive 
or nonreactive, and are  further naturally classified as direct or non-direct (roaming). 
The latter long-lived trajectories are trapped in the region of non-linear mechanical resonances, 
which in turn define alternative reaction pathways
in phase space. It is demonstrated that resonances associated with periodic orbits 
provide a dynamical explanation of the quantum mechanical resonances 
found in the isomerization rate constant calculations by Gezelter and Miller
[J.\ Chem.\ Phys.\  {\bf 103}, 7868-7876  (1995)]. 
Evidence of the trapping of trajectories by `sticky' resonant periodic orbits is provided by 
plotting Poincar\'e surfaces of section, 
and a gap time analysis is carried out in order to investigate the statistical assumption inherent in 
transition state theory for Ketene isomerization.
 
\end{abstract}

\section{Introduction}
\label{intro}

The photodissociation of ketene, CH$_2$CO, has been the subject of many studies, 
both experimental and theoretical. Moore and co-workers,
in a series of experiments, have reported many interesting aspects of this reaction 
\cite{lovejoy1991kinetics,lovejoy1992observation,lovejoy1993structures}. 
The dissociation gives rise to two product fragments, CO and 
CH$_2$. However, when the dissociation is observed 
in molecules for which the two carbons are differentiated by using different isotopes, with for example, $^{12}$CH$_2$$^{13}$CO 
or $^{13}$CH$_2$$^{12}$CO, then the products of the dissociation show in each case a mixture of $^{13}$CO and $^{12}$CO with co-fragments 
$^{13}$CH$_2$ and $^{12}$CH$_2$. These results show that isomerization 
(carbon exchange) in ketene has taken place prior to dissociation. 
The postulated mechanism involved in the isomerization of ketene is the Wolff rearrangement mechanism \cite{kirmse2002100}. In order 
to examine this hypothesis, \textit{ab initio} calculations have been carried out by several authors \cite{lovejoy1991kinetics,scott1994wolff}. 
These calculations showed that the relevant portion of the potential energy surface (PES) 
for ketene isomerization has three different 
minima associated with two symmetrically related formylmethylene species 
and an oxirene structure situated midway between these structures, respectively \cite{lovejoy1993structures}. 
On each side a high barrier 
leads to the two isomers of ketene. These findings point out the importance 
of understanding the isomerization of ketene when studying its dissociation.

Recently, Ulusoy  \textit{et al}. \cite{ulusoy2013effects,Ulusoy13b} studied the effect of roaming trajectories on the reaction rates for the
isomerization of ketene. The  {\em roaming phenomenon} is a recently identified 
chemical reaction mechanism which has attracted 
much attention due to its unusual character. 
The roaming phenomenon
was discovered
in the photodissociation of H$_2$CO \cite{townsend2004roaming}. In this process, the formaldehyde molecule can dissociate via two channels: H$_2$CO $\rightarrow$ 
H + HCO (radical channel) and H$_2$CO $\rightarrow$ H$_2$ + CO (molecular channel). Zee \textit{et al}. \cite{van1993evidence} pointed 
out that, above the threshold for the H + HCO dissociation channel, the CO rotational state distribution exhibited an intriguing ``shoulder'' at lower 
rotational levels correlated with a hot vibrational distribution of H$_2$ co-product. The observed product state distribution did not fit well with the 
traditional picture of the dissociation of formaldehyde via a well characterized saddle point transition state for the molecular channel. The roaming 
mechanism as an alternative reaction pathway, which explains the observations of van Zee and co-workers, 
was demonstrated both experimentally and in trajectory simulations by Townsend 
\textit{et al}. \cite{townsend2004roaming}. Following this work, roaming has been identified in the unimolecular dissociation of molecules 
such as CH$_3$CHO, CH$_3$OOH or CH$_3$CCH, as well as in ion-molecule reactions \cite{yu2011ab}, and it is now recognized 
as a general phenomenon in reaction dynamics (see Ref.\cite{bowman2011roaming} and references therein).

A common characteristic of systems exhibiting roaming reactions studied so far is the presence of long range interactions between the fragments 
of the dissociating molecule. However, the study by Ulusoy \textit{et al.} \cite{ulusoy2013effects,Ulusoy13b} discussed the roaming phenomenon in a different 
context. For Ketene isomerization there are no long range interactions at play 
and this raises the question of the broader relevance of the roaming mechanism.
Ulusoy \textit{et al.} \cite{ulusoy2013effects,Ulusoy13b} in their effort to find trajectories that avoid the minimum energy path (MEP) on the 
potential energy surface carried out isomerization rate constant calculations at very high energies, accessible neither to 
experiments \cite{lovejoy1991kinetics,lovejoy1992observation,lovejoy1993structures} nor to the 
quantum mechanical calculations of Gezelter and Miller \cite{gezelter1995resonant}.

In two recent studies \cite{Mauguiere2014282,MauguiereRoaming} 
we have analysed the roaming phenomenon for ion-molecule reactions.
In such reactions, the long-range 
potential function is dominated by charge induced dipole interactions; a useful model 
for these systems is  the Chesnavich potential
\cite{Chesnavich1986}. By investigating the dynamics in its appropriate setting, phase space, 
we examined the roaming phenomenon in the presence of well defined dividing surfaces (DSs) and associated reaction pathways in phase space. 
This approach enabled us to interpret roaming as a trapping phenomenon of trajectories between two DSs and the enhancement of this trapping
by the presence of non-linear mechanical resonances between the different degrees of freedom (DoF) of the system. 
Ketene offers the opportunity to investigate the roaming phenomenon in other type of molecules than those studied up to now and
this is the main purpose of the present article.

The paper is organised as follows. 
Section~\ref{sec:2} presents the reduced dimensional model for ketene isomerization used in our study and originally
introduced by Gezelter and Miller \cite{gezelter1995resonant}. 
We then analyse the dynamics of ketene isomerization in Section~\ref{sec:3}. 
In this section we first discuss the construction of DSs that dictate the reactions studied in Subsection~\ref{sec:3.1}, 
and then the dynamics in Subsection~\ref{sec:3.2} .
The next subsection, \ref{sec:3.3}, investigates the statistical assumption for the dynamics by a gap time analysis. 
Subsection~\ref{sec:3.4} presents a detailed analysis of the trapping mechanism of trajectories between the DSs and 
the role of the ubiquitous resonances due to bifurcations of periodic orbits. 
In Section~\ref{sec:4}, roaming observed in the two distinctly different systems, ion-molecule and ketene, 
is compared.
By associating roaming to the phase space structure of the molecule in a dynamically well-defined fashion, 
we establish its general character as a framework for 
revealing and exploring new reaction pathways. 
Finally, Section~\ref{sec:5} concludes.

\section{Reduced dimensional model of ketene isomerization}
\label{sec:2}

In this section we give the details of the Hamiltonian used in our study. 
A reduced dimensional model for the study of Ketene isomerization 
was proposed several years ago by Gezelter and Miller \cite{gezelter1995resonant}. 
In their original paper, they proposed
three models for ketene isomerization, retaining one, two and three relevant DoF for 
describing the isomerization
process of ketene among the nine DoF of ketene molecule, respectively. 
They subsequently employed these three models for calculating quantum 
mechanical isomerization reaction rates. From their study it appears that a 
two dimensional model provides the best agreement when compared with 
experimental results by Lovejoy et \textit{al.} \cite{lovejoy1993structures}. 
Ulusoy \textit{et al.} employed this two DoF 
model in their classical mechanical study \cite{ulusoy2013effects,Ulusoy13b} 
of the dynamics of isomerization of ketene, and we use the same model in the present study.

In the Gezelter and Miller model, one of two DoF is identified with the reactive mode and involves mainly oxygen 
and hydrogen atom motion. The other relevant DoF corresponds to the out-of-plane motion of a hydrogen atom relative to the CCO plane, 
a motion which has a harmonic frequency of approximately 514 cm$^{-1}$. Gezelter and Miller used an analytical form for their reduced dimensional
potential having several adjustable parameters in it. They fitted these parameters to reproduce the data from \textit{ab initio} calculations obtained
by Scott \textit{et al.} \cite{scott1994wolff}. The resulting potential\footnote{Note that, the original paper by Gezelter and Miller \cite{gezelter1995resonant} had a typo in it resulting in a missing factor of
one half in front of the second term in the RHS of their equation (2.2).}  is expressed as:

\begin{equation}
V(q_F,q_1) = V_{1d}(q_F) + \frac{1}{2} k \left( q_1+\frac{d}{k}\,q_F^4 \right)^2,
\label{eq:1}
\end{equation}
with

\begin{equation}
V_{1d}(q_F) = a_2 \, q_F^2 + a_4 \, q_F^4 + a_6 \, q_F^6 + c \, q_F^2 \exp(-d_0 \, q_F^2).
\label{eq:2}
\end{equation}
Numerical values of the potential parameters are given in  Table \ref{tab:1}.
Fig~\ref{fig:1} depicts a contour plot of the
potential along with some periodic orbits (see Subsection~\ref{sec:3.4}). The different equilibrium points (EP) of the potential
are listed in Table~\ref{tab:2}. For each EP the stability is indicated by the labels CC or CS. A CC label denotes
a centre-centre EP, which means that the EP is stable in both directions and the label CS is used for
center-saddle, which means that the EP is stable in one direction and unstable in the other.

The total Hamiltonian is obtained by adding to the above potential a kinetic energy term quadratic
in momenta ($p_{q_F}$,$p_{q_1}$) conjugate to the coordinates ($q_F$,$q_1$), respectively. The resulting Hamiltonian
takes the form:

\begin{equation}
H(p_{q_F},p_{q_1},q_F,q_1) = \frac{p^2_{q_F}}{2m_O} + \frac{p^2_{q_1}}{2m_H} + V(q_F,q_1),
\label{eq:3}
\end{equation}
where $m_O = 16 $ u and $m_H = 1 $ u are the masses of oxygen and hydrogen atom, respectively. 
This choice of masses
was used by Gezelter and Miller \cite{Gezelterprivate}.

\begin{table}
\begin{center}
\caption{Parameters used in the reduced dimensional potential of Ketene.}
\label{tab:1}       
\begin{tabular}{|lc|c|c||}
\hline
parameter & value & units \\
\hline\hline
$k$      &  $1.0074 \times 10^{-2}$  & $  (E_h a_0^2)$ \\
            &                                           &  \\
$d$      &  $1.9769$  & $ (E_h a_0^5)$ \\
            &                                           &  \\
$a_2$  &   $-2.3597 \times 10^{-3}$  & $ (E_h a_0^2)$ \\
            &                                           &  \\
$a_4$  &   $1.0408  \times10^{-3}$  & $ (E_h a_0^4)$ \\
            &                                           &  \\
$a_6$  &   $ -7.5496 \times10^{-5}$  & $ (E_h a_0^6)$ \\
            &                                           &  \\
$c$      &   $ 7.7569  \times10^{-3}$  & $ (E_h a_0^2)$ \\ 
            &                                           &  \\
$d_0$ &    $-2.45182 \times10^{-4}$  & $ (a_0^2)$ \\
\hline\hline
\end{tabular}
\end{center}
\end{table}

\begin{table}
\begin{center}
\caption{Equilibrium points of the potential with their stability. Coordinates $q_F$ and $q_1$
are indicated in Bohr. CS means center-saddle and CC centre-centre.}
\label{tab:2}       
\begin{tabular}{|lclc|c|c|}
\hline
Name & $q_F$ & $q_1$ & stability \\
\hline\hline
EP$_1$      &   -2.805  &  1.506  & CS \\
EP$_2$      &  -1.325   &  0.075  & CC \\
EP$_3$      &  -0.547   &  0.002  & CS \\
EP$_4$      &  0.0         &  0.0       & CC \\
EP$_5$      &  0.547    &  0.002  & CS \\
EP$_6$      &  1.325    &  0.075  & CC \\
EP$_7$      &   2.805   &  1.506  & CS \\
\hline\hline
\end{tabular}
\end{center}
\end{table}

\section{Dynamics of ketene isomerization}
\label{sec:3}

In this work we are concerned with the dynamics of isomerization in Ketene and its relation to
the roaming phenomenon. Recently, Ulusoy \textit{et al.} reported a study of ketene 
isomerization 
where they analysed the impact of roaming trajectories on the 
rate of the reaction \cite{ulusoy2013effects,Ulusoy13b}. Up to now roaming reactions have been identified 
in the context of reactions involving long range interactions between two fragments of a dissociating
molecule. The study of Ulusoy \textit{et al.} suggests the possibility of
looking at roaming in different reaction dynamics
contexts and in this sense represents an 
important conceptual advance. Recently, we studied the dynamics of roaming 
reactions in the context of ion-molecule reactions \cite{Mauguiere2014282,MauguiereRoaming} and proposed a dynamical origin for the 
roaming phenomenon. Our approach consisted of identifying optimal DSs for the different reactive
events possibly occurring in the problem considered. The roaming phenomenon was then identified with the trapping
of trajectories between two DSs and enhanced by the presence of non-linear resonances identified with resonant
periodic orbits (POs). In the present work we extend our methods to study the ketene isomerization reaction and the way roaming appears.
To begin, we must first locate the relevant DSs for the problem we consider. The next paragraph
is devoted to the definition of proper DSs and how this can be achieved with the help of the so-called normally hyperbolic invariant
manifolds (NHIMs).

\subsection{Normally hyperbolic invariant manifolds and dividing surfaces}
\label{sec:3.1}

The study of reaction dynamics aims at understanding the associated mechanism 
at an atomic level of detail in a specific molecular
reaction. Eventually, one is interested in computing quantities such as reaction rate constants, which 
can be compared with  experiment. The most popular theory developed for this purpose is transition state theory
(TST). TST relies on several fundamental assumptions \cite{Baer96}.
One of these assumptions is the existence of a dividing surface
which separates reactants from products and which have the non-recrossing property. This means that a trajectory
initiated on the reactants and crossing this dividing surface must proceed to products without recrossing the dividing surface, 
and end up as a reactants. The other fundamental assumption of TST is that of statistical dynamics. If one considers the reaction 
at a specific energy (microcanonical), the statistical assumption 
requires that throughout the dissociation of the molecule all phase space 
points are equally probable on the time scale of reaction \cite{Baer96}. 
This assumption is equivalent to saying that the 
redistribution of the energy amongst the different DoF of the system on the reactant side of the DS is fast compared to the 
rate of the reaction, and this guarantees a single exponential decay for the reaction. 
We will check this assumption for the ketene model by a gap time analysis in the roaming region (see below).

In our previous study of roaming phenomenon, we explained how NHIMs provide a 
solution to the problem of the non-recrossing
property of the DS. In fact, the problem of constructing a non-recrossing DS for a two DoF system was solved 
during the 1970s by Pechukas, Pollak and coworkers \cite{Pechukas73,Pechukas77,Pollak78,Pechukas79}. 
They showed that the DS 
at a specific energy is intimately related to an invariant phase space object, an unstable PO. 
The periodic orbit defines the bottleneck in phase space through which the reaction occurs 
and the DS which intersects trajectories evolving from reactants to products can be shown to have the topology of a hemisphere, whose 
boundary is the PO. As our present system is a two DoF problem, we will focus on POs and the DSs constructed from them. Nevertheless, 
this unstable PO is a simple example of a NHIM. Generalisation to higher number of DoF has been a major
obstacle in the development of the theory, and the question of the construction of
DSs for systems with three and more DoF has been given a satisfactory answer only recently. 
The appropriate construction is built around an invariant phase space object,
a NHIM, which is the generalisation of the unstable PO of the two DoF case. 
The NHIM serves as the anchor for the construction
of the DS and this DS constructed in this way can be shown to have the non-recrossing property \cite{waalkens2004direct}.

\begin{figure}
\includegraphics[scale=0.8]{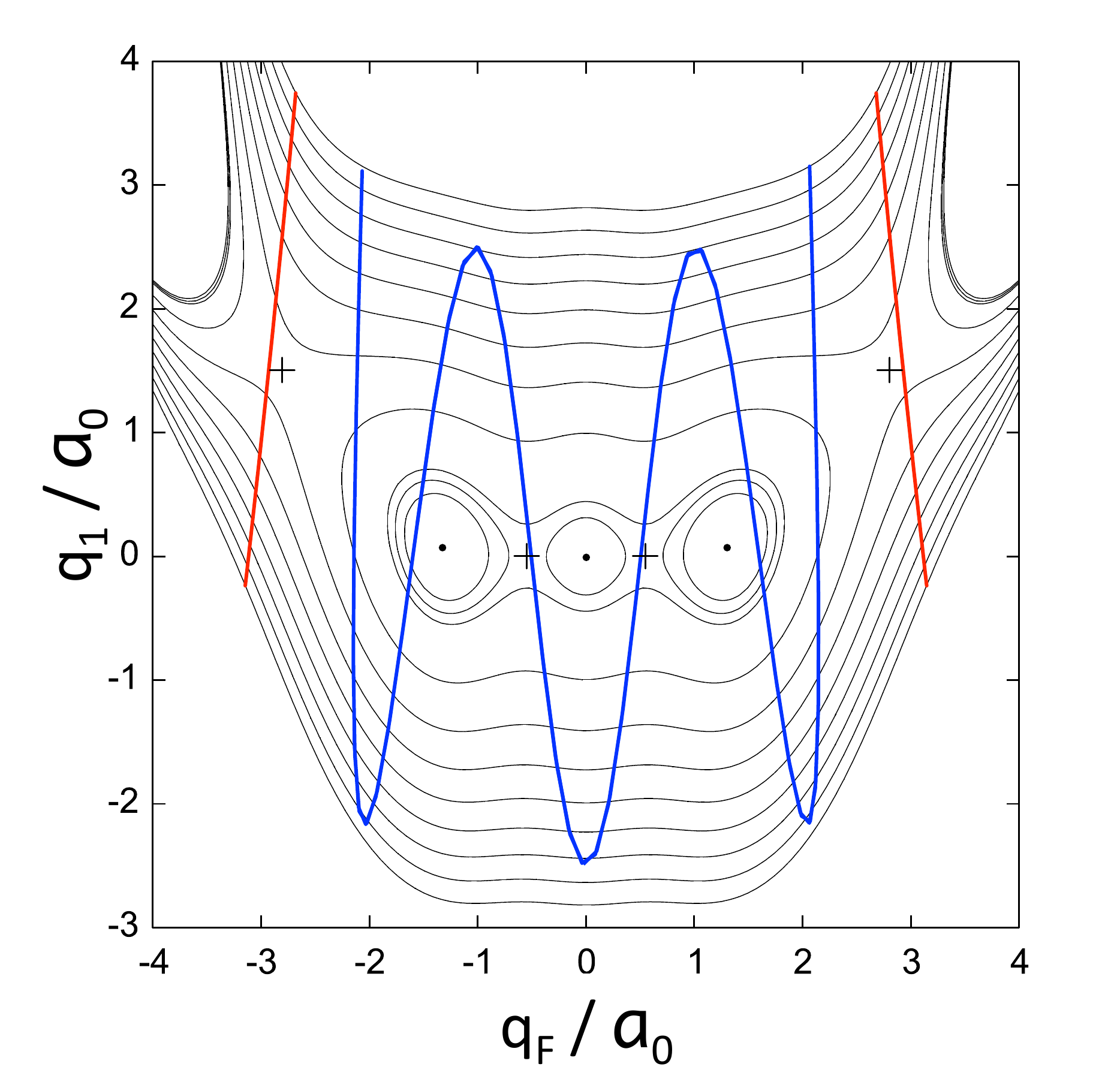}
\caption{Contour plot of the PES along with some POs. The red curves are the Lyapunov POs and the blue is a 1:6 resonant PO (see text).}
\label{fig:1} 
\end{figure}

The construction of a non-recrossing DS for the simple two DoF model of ketene isomerization 
therefore starts with the location
of relevant unstable POs. 
For the ketene problem isomerization reaction is completed
when the system passes from one Ketene well to the other. 
The model we are dealing with locates the oxirene minimum
between the formylmethylene wells. 
If we consider that the system is initially in the ketene well located on the part of the
potential for which $q_F<-3\; {a}_0$ (the ``left'' ketene well if we look at Fig~\ref{fig:1}), then 
isomerization is completed 
if the system passes to the other ketene well located on the part of the potential 
for which $q_F>3 \; {a}_0$. 
To accomplish this reaction we see that the system must pass through two bottlenecks located around
the equilibria labelled EP$_1$ and EP$_7$ in Table~\ref{tab:2}. These two EPs are of CS stability type, which means that the 
linearized vector field at these points (with momenta $p_{q_F}=0$ and $p_{q_1}=0$) has one pair of real eigenvalues and one
pair of imaginary eigenvalues. This means that in the vicinity of these EPs there exist unstable POs \cite{Moser76}, 
the so-called Lyapunov POs \cite{Kelley1969}.
In the vicinity of these EPs, the Lyapunov POs exist as families of POs depending on the energy of the system. 
Respresentative POs of these families are shown in  Fig~\ref{fig:1}.
These POs are the NHIMs which define the bottlenecks through which the reaction occurs and will serve as anchors for the construction
of the non-recrossing DSs.

\begin{figure}
\includegraphics[scale=0.8]{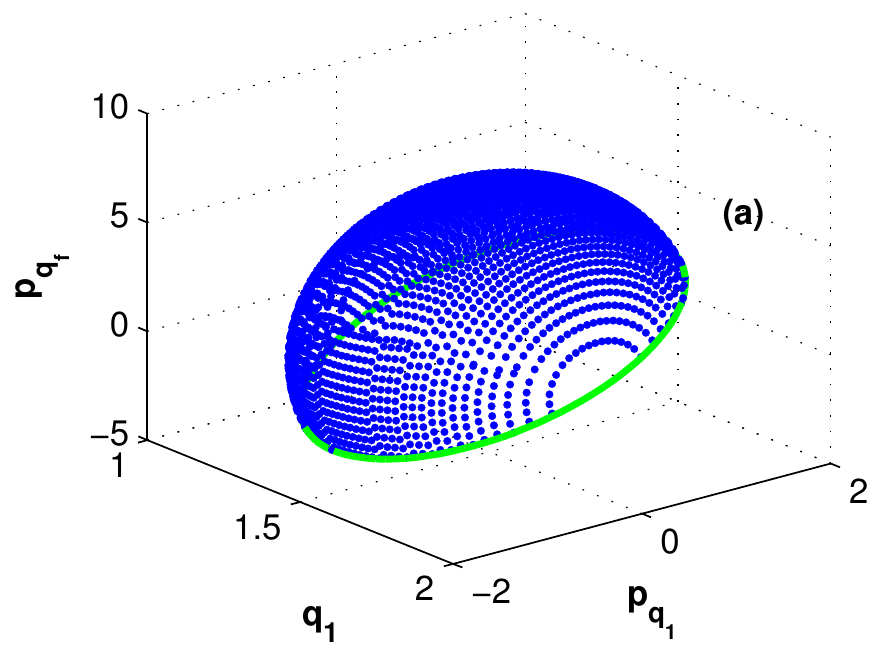}
\includegraphics[scale=0.8]{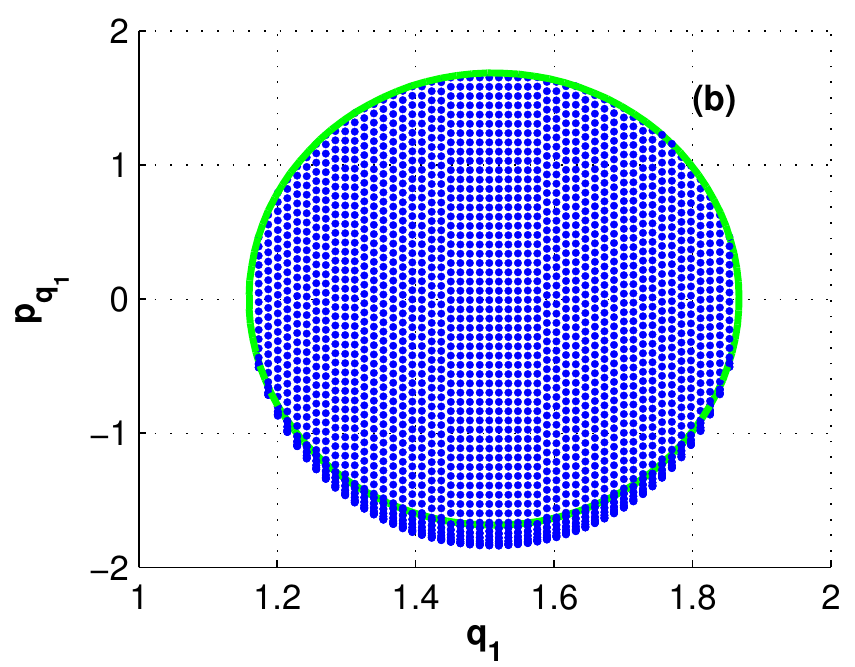}
\caption{Forward hemisphere of the dividing surface at the top left saddle at energy $E=0.01$ $E_h$. (a) 3-D view. (b) projection on the ($q_1,p_{q_1}$) plane.}
\label{fig:2} 
\end{figure}

The DSs attached to these NHIMs are obtained by considering a fixed value of energy. At each energy two Lyapunov POs exist,
each located in the vicinity of EP$_1$ and EP$_7$, respectively. These POs are the equators of the DSs \cite{waalkens2004direct} 
and split the DSs into two hemispheres. One hemisphere intersects trajectories travelling from reactant to products (forward hemisphere) 
and the other intersects trajectories travelling from products to reactants (backward hemisphere). For concreteness, 
we consider the Lyapunov PO at a certain energy in the vicinity of EP$_1$. This PO projects
to a line in configuration space (see Fig~\ref{fig:1}). To sample points on the DS we need to specify the phase space coordinates of the points
belonging to the DS at a certain energy. To do so, we move along the PO (the equator of the DS) in configuration space. This fixes the values of
$q_F$ and $q_1$. We then scan the values of one of the momenta $p_{q_F}$ or $p_{q_1}$. To obtain the value of the remaining phase space
coordinate (the remaining momentum) we solve the equation $H(p_{q_F},p_{q_1},q_F,q_1)=E$ for the remaining unknown 
momentum ($p_{q_F}$ or $p_{q_1}$) and at a specified value of energy, $E$. Since, the kinetic energy is quadratic in the momenta, 
this equation has two real roots distinguished by their sign.
Each root belongs to one of the hemispheres of the DS (forward and backward hemispheres). The result of this sampling procedure is shown in 
Fig~\ref{fig:2}. Panel (a) of this figure is a 3D plot of the forward hemisphere (blue dots) of the DS at energy of 0.01 Hartree 
above the oxirene minimum. The Lyapunov PO is shown by the green curve.
Panel (b) depicts a 2D projection of this hemisphere onto the ($q_1$,$p_{q_1}$) plane. At the bottom of this figure we note that some points
belonging to the hemisphere lie out of the area enclosed by the Lyapunov PO (green curve). 
This is the result of the `tilting' over of the DS in the
coordinates we are using. Because of this fact, to obtain a uniform sampling with 
full coverage of the hemisphere we sample momenta 
in two ways: sampling $p_{q_1}$ with $p_{q_F}$ obtained by solving the energy equation mentioned above,
or sampling $p_{q_F}$ and fixing $p_{q_1}$ 
from the equation of energy.

\subsection{Classical trajectory simulations}
\label{sec:3.2}

Our aim is to investigate the dynamics of ketene isomerization and how one can 
identify and understand the roaming phenomenon in
this situation. To do so, we propagated classical trajectories and examined 
their qualitatively different characters.

If we consider that the system is initially located in the ketene well 
in the region where $q_F<-3\; {a}_0$, then the isomerization
reaction is completed when the system arrives in the ketene well on the region where $q_F>3 \; {a}_0$. 
On its way to complete this reaction,
a trajectory has to pass through two bottlenecks, one located in the vicinity of EP$_1$ and which is defined by the Lyapunov PO
we discussed above, and the other located in the vicinity of EP$_7$ also defined by a Lyapunov PO. We explained above how to
construct DSs attached to these two POs. In order to understand the dynamics of the isomerization process we sampled the DS attached 
to the Lyapunov PO associated with EP$_1$. We used these samples as initial conditions for numerically propagating  classical trajectories.

As we did in our previous studies \cite{Mauguiere2014282,MauguiereRoaming} of roaming phenomenon we classified the trajectories into
four qualitatively different categories. 
These different categories of trajectories are defined 
in relation to the reactive scenario involved in the ketene isomerization. 
To begin, we note that there are two obvious different types of expected trajectories. 
First, there are \textit{reactive}  trajectories which will complete the isomerization reaction, 
which means that these type of trajectories will cross the DS located in the vicinity of EP$_7$.
Second, there are \textit{non-reactive} trajectories which do not complete the isomerization reaction, but instead, recross the DS located in the vicinity of EP$_1$ 
and end up in the ketene well where they originally came from. 

Since we are interested in understanding the roaming phenomenon 
from a dynamical point of view, we need to specify
those characteristics a trajectory should have to be assigned as roaming. 
The notion of roaming is invoked when a trajectory does not follow  the minimum energy path (MEP) or 
intrinsic reaction coordinate (IRC) \cite{heidrich1995reaction} of the potential function. 
In our problem this departure from traditional understanding of reactivity
occurs in trajectories which do not
complete \textit{directly} the isomerization reaction or do not recross \textit{directly} 
the DS located in the vicinity of EP$_1$. 
Instead, the \textit{roaming} trajectories spend more time in the trapping (roaming) region and 
exhibit oscillations in the $q_F$ direction before eventually finding their way out of the oxirene well region 
by either crossing the DS located in the vicinity of EP$_7$ or the DS located in the vicinity of EP$_1$. 
To summarise, reactive (which complete the isomerization reaction) 
and non-reactive (which do not complete isomerization) trajectories 
are further categorized as direct and roaming. 
Hence, we have four classes of trajectories: direct reactive, roaming reactive,
direct non-reactive and roaming non-reactive. 
However, in order to quantitatively define a roaming trajectory, 
we need a criterion to decide when a trajectory exhibit oscillations in the $q_F$
direction before eventually becoming reacting or non-reacting. 
To achieve this task we count the number of times a trajectory crosses 
the symmetry line $q_F=0$. The precise definitions of  the four categories are therefore:

\begin{description}
\item [1)] Direct reactive trajectories: these trajectories cross  the line $q_F=0$ 
only once before crossing the DS located in the vicinity of EP$_7$ and completing the isomerization reaction.

\item [2)] Roaming reactive trajectories: these trajectories cross the line $q_F=0$
at least three times  before crossing the DS located in the vicinity of EP$_7$ and complete the isomerization reaction. 
Note that a reactive trajectory has to cross the line $q_F=0$ an odd number of times.

\item [3)] Direct non-reactive trajectories: these trajectories cross  the line $q_F=0$ 
only twice before crossing the DS located in the vicinity of EP$_1$, and do \textit{not} 
complete the isomerization reaction.

\item [4)] Roaming non-reactive trajectories: these trajectories cross  the line $q_F=0$
at least four times before crossing the DS located in the vicinity of EP$_1$ and do \textit{not} complete the isomerization reaction. 
Note that non-reactive trajectories have to cross the line $q_F=0$ an even number of times.

\end{description}

\begin{figure}
\includegraphics[scale=0.8]{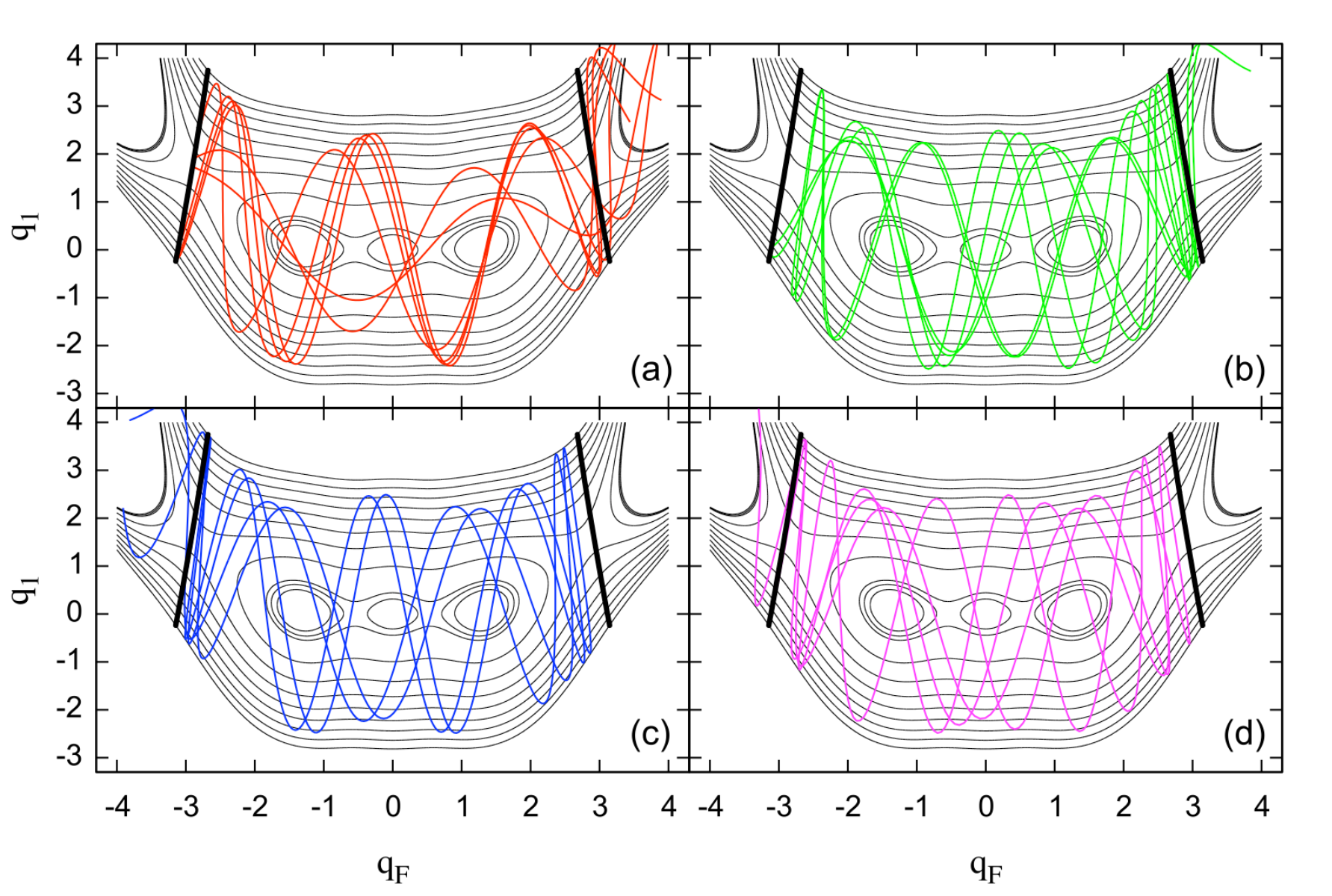}
\caption{The four different categories of trajectories at energy $E=0.04$ $E_h$. (a) Direct reactive trajectories. (b) Roaming reactive trajectories.
(c) Direct non-reactive trajectories. (d) Roaming non-reactive trajectories.}
\label{fig:3} 
\end{figure}

In Fig~\ref{fig:3} we show representative trajectories from these four categories. 
Each panel in this figure corresponds to a particular category.
Panel (a) shows representative direct reactive trajectories, panel (b) roaming reactive, 
panel (c) direct non-reactive and panel (d) roaming non-reactive.

At this point, it is interesting to compare the present results on
ketene isomerization with those obtained in our
previous studies of the roaming phenomenon \cite{Mauguiere2014282,MauguiereRoaming}, which were done in the context of an ion-molecule reaction. We see that in the present problem the natural classification into four different categories of the trajectories fits nicely into the classification provided
in the previous studies. Our earlier analyses of roaming phenomenon pointed out that 
roaming is related to the appearance of trapped trajectories between two DSs in phase space. 
This led us to define a roaming region which is the region of
phase space between these two DSs, where the roaming occurs. The scenario here for ketene 
isomerization is similar and the trapping occurs 
between the two DSs located in the vicinity of EP$_1$ and EP$_7$. 
These two DSs define a region in phase space which can be identified with the roaming 
region we defined in our previous work. Furthermore, we previously described how non-linear resonances 
manifest by the presence of certain POs provide the basic mechanism 
to transfer energy from one DoF to another 
and enhance the trapping phenomenon 
in the roaming region. We shall see below that this ingredient of the 
roaming phenomenon is also present in 
ketene isomerization and takes also the form of resonant POs. 
Lastly, we noted in our earlier studies that our approach had strong 
similarities with Miller's unified approach to reaction mechanisms in the presence
of complex formation \cite{Miller76a}. 
Specifically, Miller considered how TST should be modified in the presence
of a complex formation. This led us to investigate the question of 
the statistical assumption of TST in the roaming region using
a gap time analysis, and we now investigate this question for 
the ketene isomerization.

\subsection{Gap time analysis}
\label{sec:3.3}

The relevance of studying gap time in unimolecular reaction
is seen in the investigation of the validity of the statistical assumption of TST. 
This assumption has been the subject of numerous works
(see, for example the work of Slater \cite{Slater56,Slater59}, Bunker \cite{Bunker62,Bunker64}, Bunker and Hase \cite{Bunker73},
Thiele \cite{Thiele62,Thiele62a}, Dumont and Brumer \cite{Dumont86} 
and DeLeon and co-workers \cite{DeLeon81,Berne82}).

In our earlier paper \cite{MauguiereRoaming} we summarised the gap time approach 
to reaction rates due to Thiele \cite{Thiele62,Thiele62a} (see also 
the discussion in Ref.~\cite{ezra2009microcanonical}). 
The gap time in the  ketene isomerization problem is the time $s$ it 
takes for a trajectory to traverse the roaming region. 
If we consider a point on the forward hemisphere of the DS located in the vicinity of EP$_1$ 
(the hemisphere which intercepts trajectories travelling from the ``left'' Ketene well to the oxirene well), 
the gap time of a trajectory initiated at this point is the time it takes for this trajectory
to reach either the hemisphere intercepting trajectories travelling from the oxirene well to the ``right'' ketene well of the DS located in the vicinity of
EP$_7$ or to reach the backward hemisphere of the DS located in the vicinity of EP$_1$ (the one which intercept trajectories travelling 
from the oxirene well to the ``left'' ketene well). An important notion in the gap time formulation of TST is the gap time distribution, $P(s;E)$: 
the probability that a phase space point on the forward hemisphere of the DS at EP$_1$ at energy $E$ has a gap time between $s$ and $s + ds$ 
is equal to $P (s; E) ds$. The statistical assumption of TST is equivalent to the requirement that the gap time distribution is the random, 
exponential distribution.

\begin{equation}
P(s; E) = k(E) \exp(-k(E)s).
\label{eq:4}
\end{equation}
This distribution is characterised by a single exponential decay constant $k(E)$, which is a function of the energy.

In addition to the gap time distribution, we also consider the integrated gap time distribution $F(t;E)$, which is defined as 
the fraction of trajectories with gap times $s \geq t$, and is expressed as

\begin{equation}
\label{eq:5}
F(t;E)  = \int_t^{+\infty} ds \; P(s;E).
\end{equation}
For the random gap time distribution the integrated gap time distribution is exponential, $F(t;E)=\exp(-kt)$.

\begin{figure}
\includegraphics[scale=0.6]{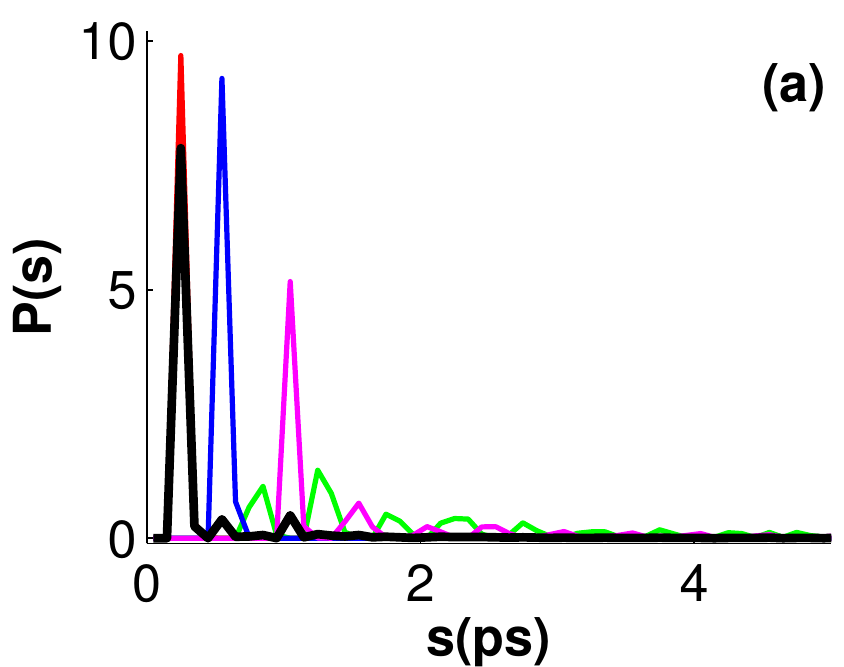}
\includegraphics[scale=0.6]{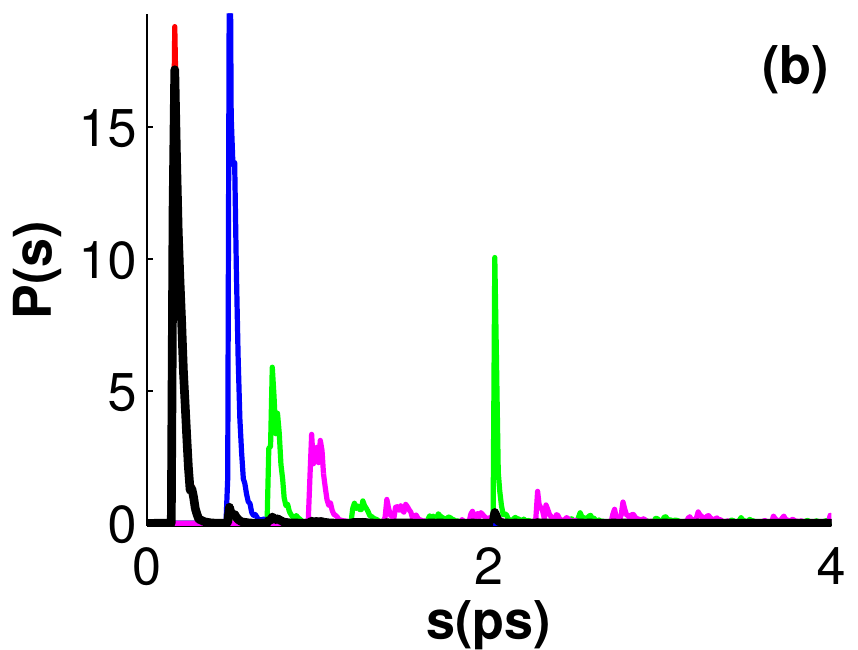}
\includegraphics[scale=0.6]{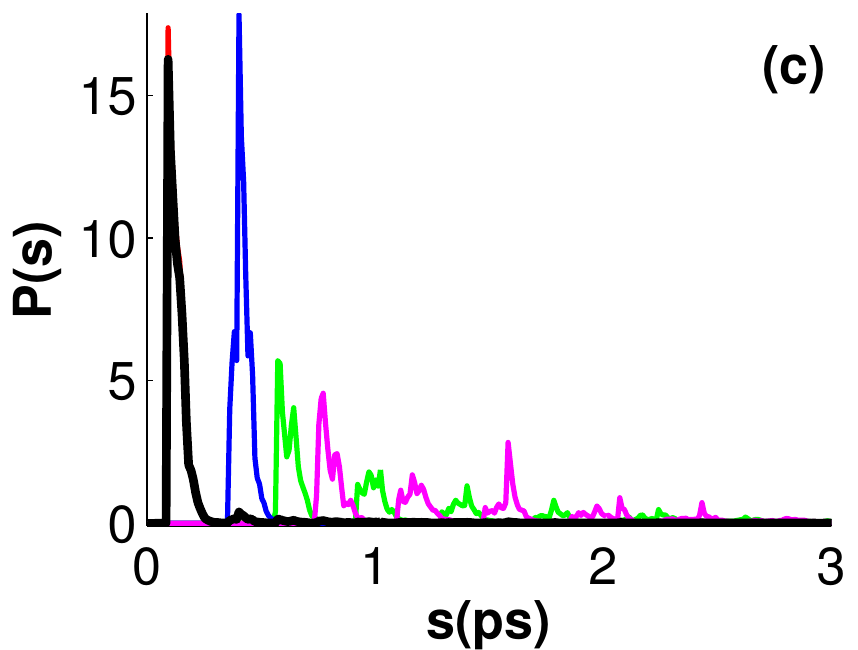}
\includegraphics[scale=0.6]{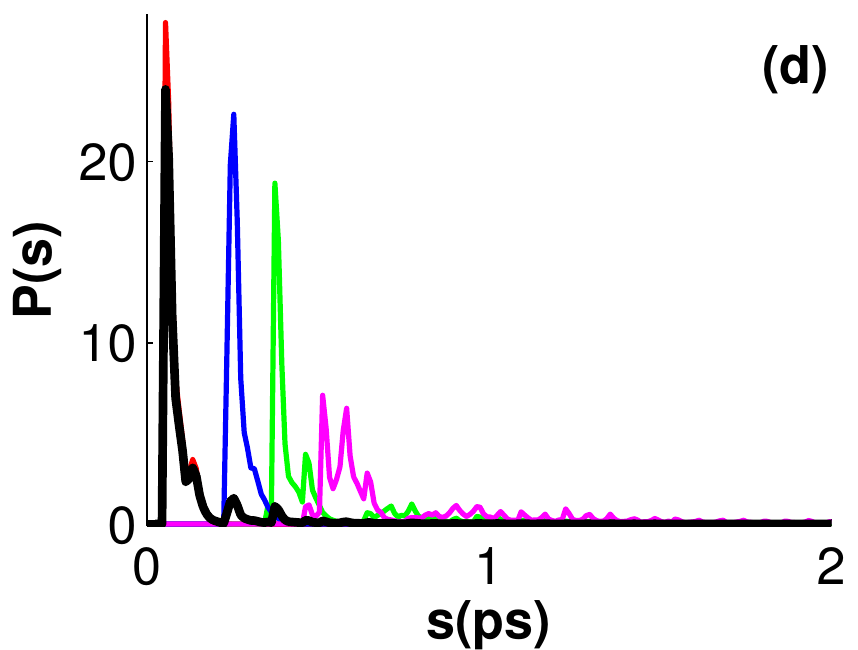}
\caption{Gap time distributions. For each energy we give the number $n$ of trajectories used to compute the distributions. (a) Energy $E=0.010$ $E_h$, $n=942592$. (b) $E=0.0150$ $E_h$, $n=941938$. (c) $E=0.040$ $E_h$, $n=1967165$. (d) $E=0.150$ $E_h$, $n=1932961$. }
\label{fig:4} 
\end{figure}

\begin{figure}
\includegraphics[scale=0.6]{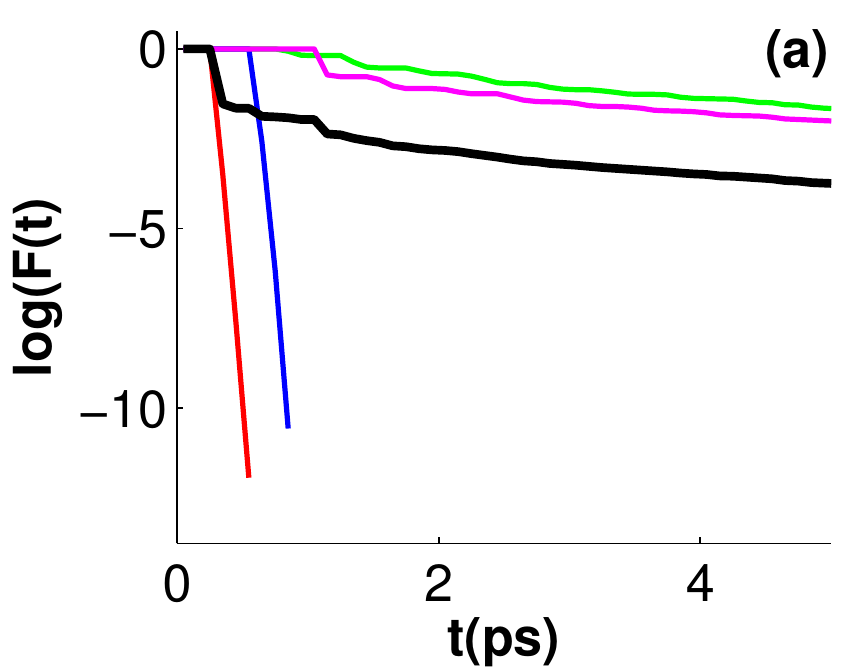}
\includegraphics[scale=0.6]{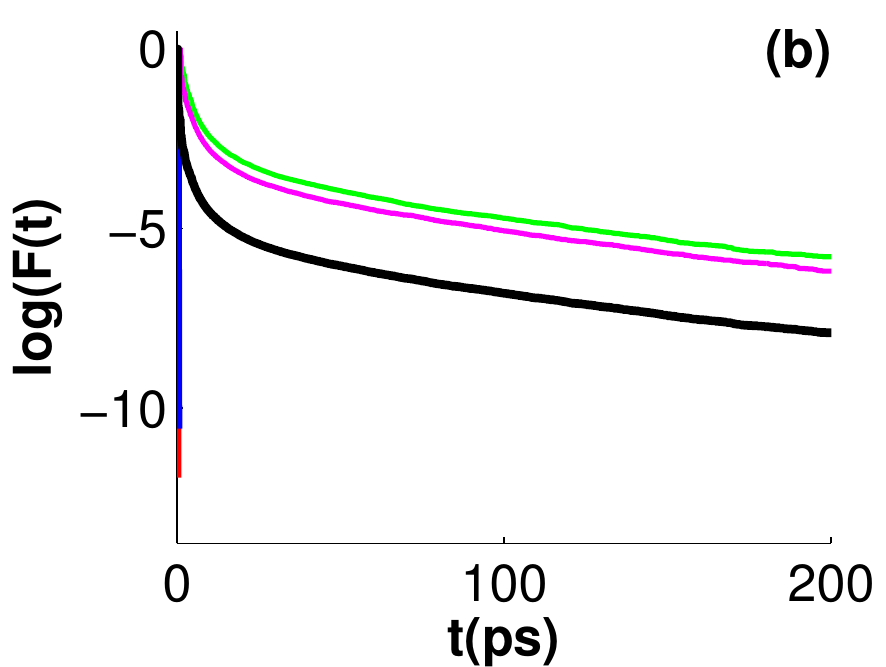}
\includegraphics[scale=0.6]{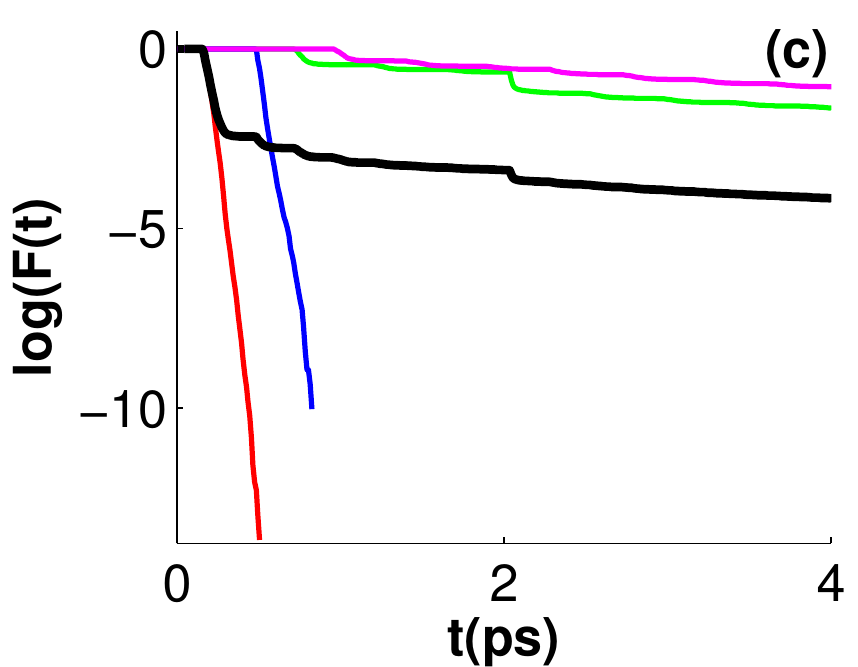}
\includegraphics[scale=0.6]{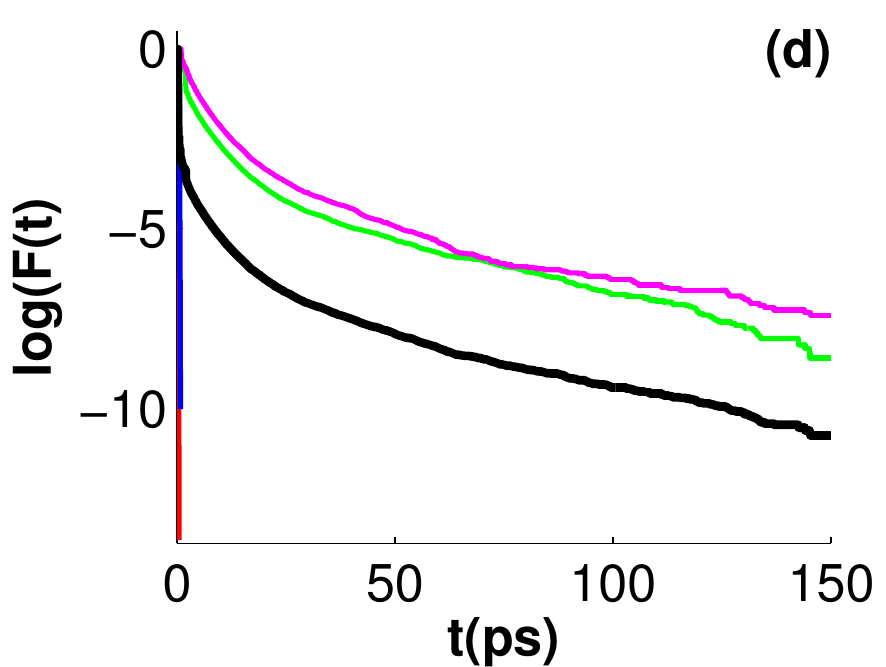}
\includegraphics[scale=0.6]{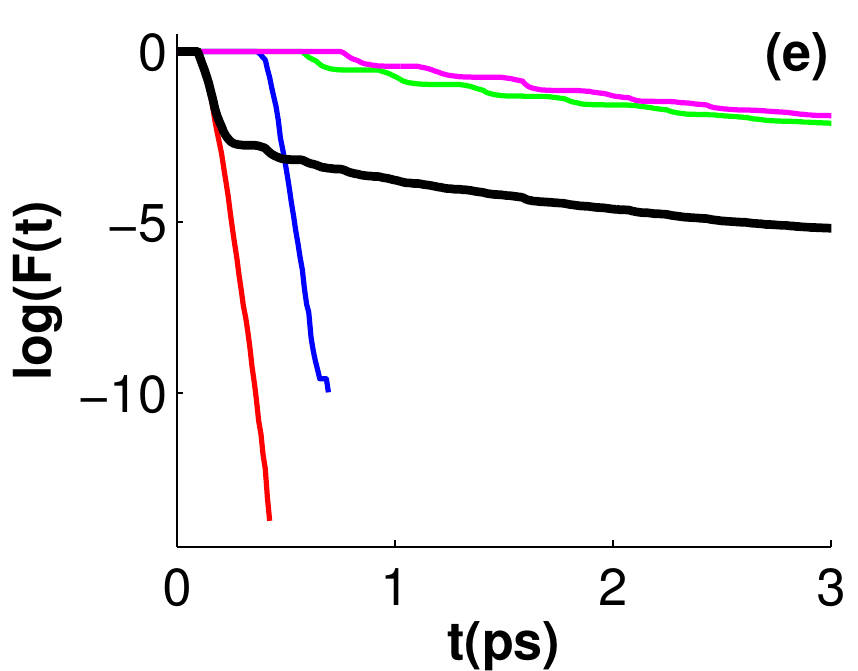}
\includegraphics[scale=0.6]{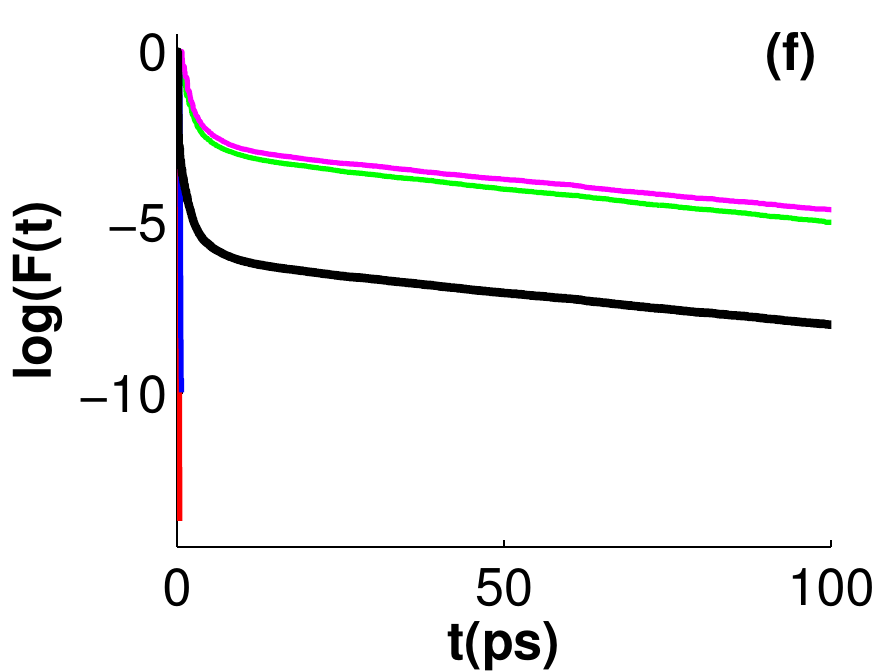}
\includegraphics[scale=0.6]{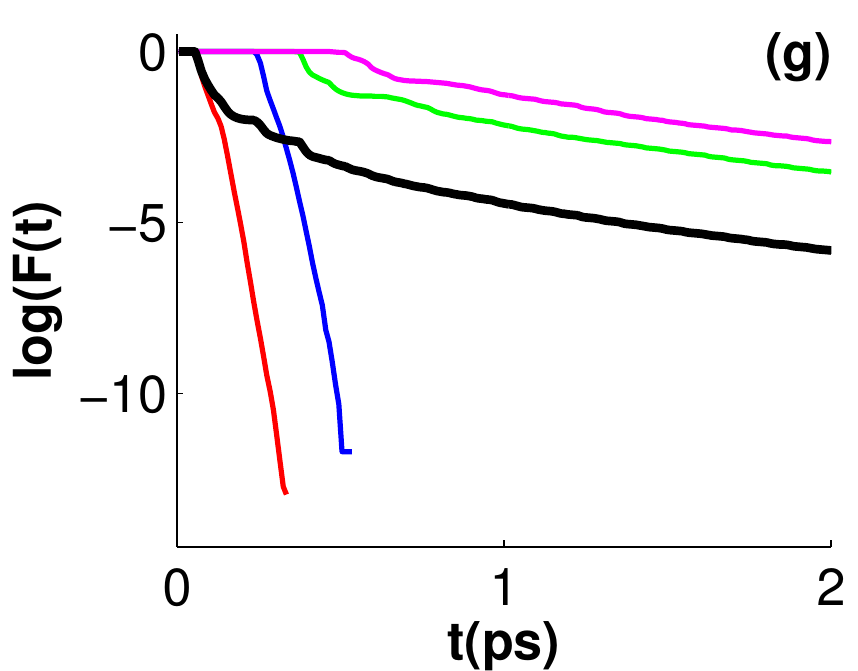}
\includegraphics[scale=0.6]{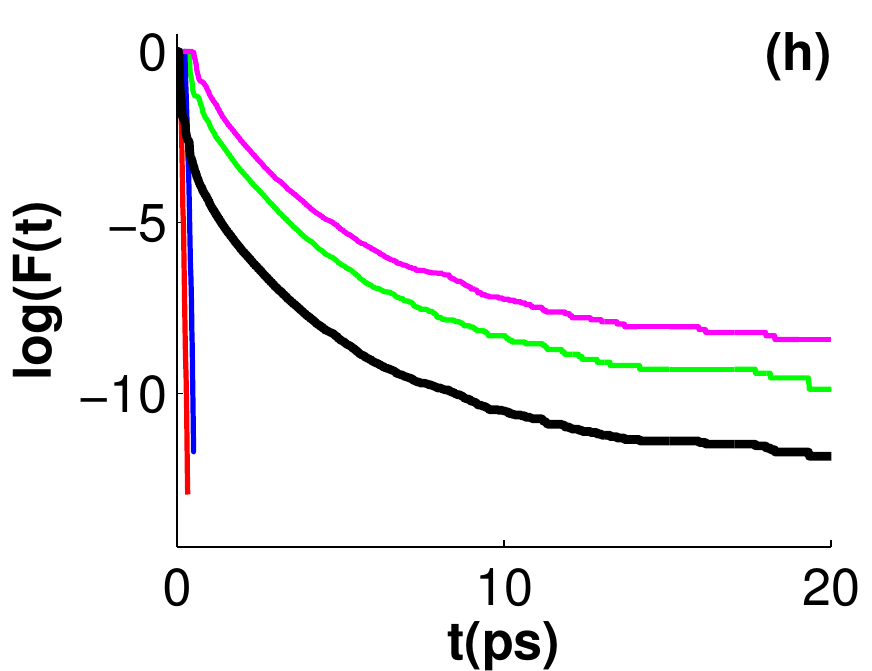}
\caption{Life time distributions. For each energy we plot two panels showing the small time scale and the large time scale of the distribution. The number of trajectories used to compute the distributions is as indicated in Fig~\ref{fig:4}. 
(a) and (b) Energy $E=0.010$ $E_h$. (c) and (d) $E=0.0150$ $E_h$. (e) and (f) $E=0.040$ $E_h$. (g) and (h) Energy $E= 0.150$ $E_h$. }
\label{fig:5} 
\end{figure}

The results of gap time and integrated gap time distributions for ketene isomerization are shown in Fig~\ref{fig:4} and Fig~\ref{fig:5}.
To obtain these figures we sampled uniformly at a certain energy the 
forward hemisphere of the DS at EP$_1$ and propagated
trajectories until they exit the roaming region by either crossing the DS at EP$_7$ or 
recrossed the DS at EP$_1$. Fig~\ref{fig:4}
shows the gap time distributions at different energies measured from the oxirene minimum (see the caption of the figure). For each panel we have
reported the individual gap time distributions of the different categories of trajectories. The red curves represent the distributions 
for the direct reactive trajectories, the green represent the roaming reactive trajectories, the blue the direct non-reactive and
the magenta the roaming non-reactive trajectories. The black curves depict the gap time distributions for all categories taken together.
Fig~\ref{fig:5} shows the integrated gap time distributions at the same energies as Fig~\ref{fig:4}. For each energy we show two panels 
presenting the short time scale distributions and the large time scale distributions. The different curves with their
colours have the same meaning as in Fig~\ref{fig:4}. 

As we can see, these two figures exhibit significant deviation from the 
random gap time distributions as well as
exponential integrated gap time distributions, indicating that 
the statistical assumption of TST is not valid for the ketene isomerization
problem and must be coorected in some fashion. To give an idea of the kind of corrections one needs to provide, 
we plot in Fig~\ref{fig:6} the fractions of the different types
of trajectories as a function of the energy. 
If one were to use the DS at EP$_1$ to compute the flux needed to evaluate the reaction
rate in TST, this flux would have to be corrected by the fractions given in Fig~\ref{fig:6}. Indeed, since some trajectories are reflected
back to the ``left'' ketene well there are some recrossing trajectories and the real flux is the flux through the DS multiplied by the sum
of the fractions of the direct and roaming reactive trajectories.

\subsection{Trapping in the roaming region and resonant periodic orbits}
\label{sec:3.4}

We have attributed the roaming phenomenon to trapped trajectories, 
which follow alternative pathways to those associated with  
MEPs on the PES. The trapping is due to multiple dividing surfaces and the existence of non-linear mechanical
resonances. Resonances 
can be traced in phase space by constructing continuation/bifurcation  diagrams of periodic orbits
starting from principal POs originating from equilibria, minima and saddles, 
and following them in a parameter space,
which may include the total energy, angular momentum, masses, or other parameters of interest. 
This is a well known strategy in studying non-linear dynamical systems. 

For the ketene isomerization model we located families of POs that 
are related to EP1 and EP7 equilibria and 
the roaming region at low energies of excitation. These families emanate from a cascade of center-saddle bifurcations 
approaching the saddle from energies above the isomerization threshold \cite{Farantos2005_110}. The results are shown in a continuation/bifurcation diagram of POs in Fig~\ref{fig:7}. 
This figure depicts the period of the POs found as a function of energy. We also label the POs with respect
to their resonance numbers $n:m$, which means that the PO makes $n$ oscillations 
in $q_F$ direction in the time it performs 
$m$ oscillations in $q_1$ direction. To give an idea of the shape of these POs in configuration space 
we have plotted the $1:6$ resonant 
PO as a blue line in Fig~\ref{fig:1}. It is easy to recognize the impact of this family of POs on the dynamics of ketene from 
the plots of representative trajectories in Fig~\ref{fig:3}. It is worth emphasizing that these periodic orbits are the results of center-saddle bifurcations. Thus, we expect, and indeed find, stable and unstable branches of POs to appear simultaneously and these periodic orbits define the
resonance region in phase space. 
A linear stability analysis of periodic orbits reveal the stable/unstable character of nearby trajectories with respect to POs.
However, in order to better estimate the size of the resonance zone defined by these periodic orbits we calculate 
Poincar\'e surfaces of section  (PSS) \cite{Wiggins_book03}, and  autocorrelation functions from batches of trajectories sampled 
around the POs.

\subsubsection{Poncar\'e surfaces of section}
\label{PSS}

\begin{figure}
\includegraphics[scale=0.3]{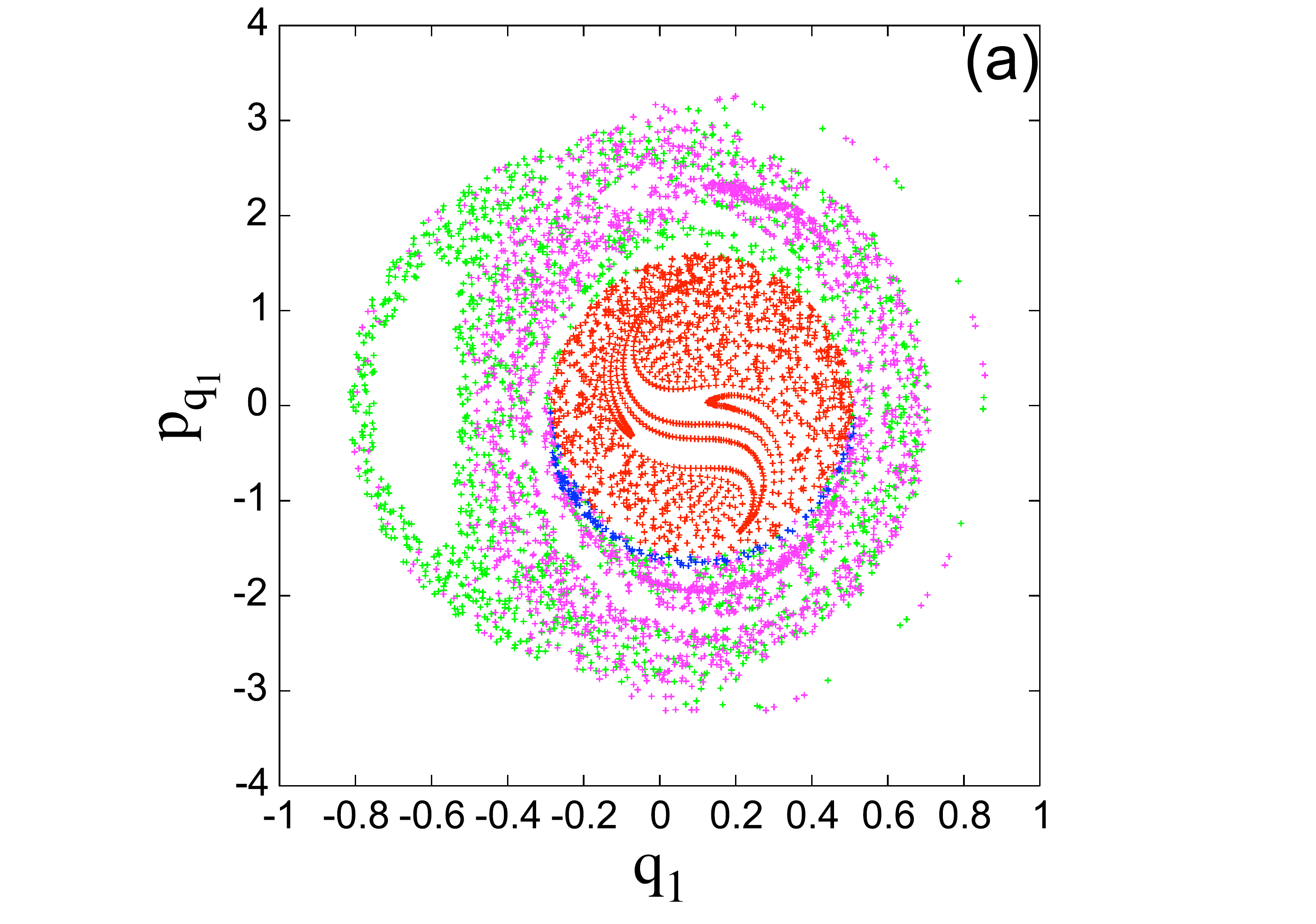}
\includegraphics[scale=0.3]{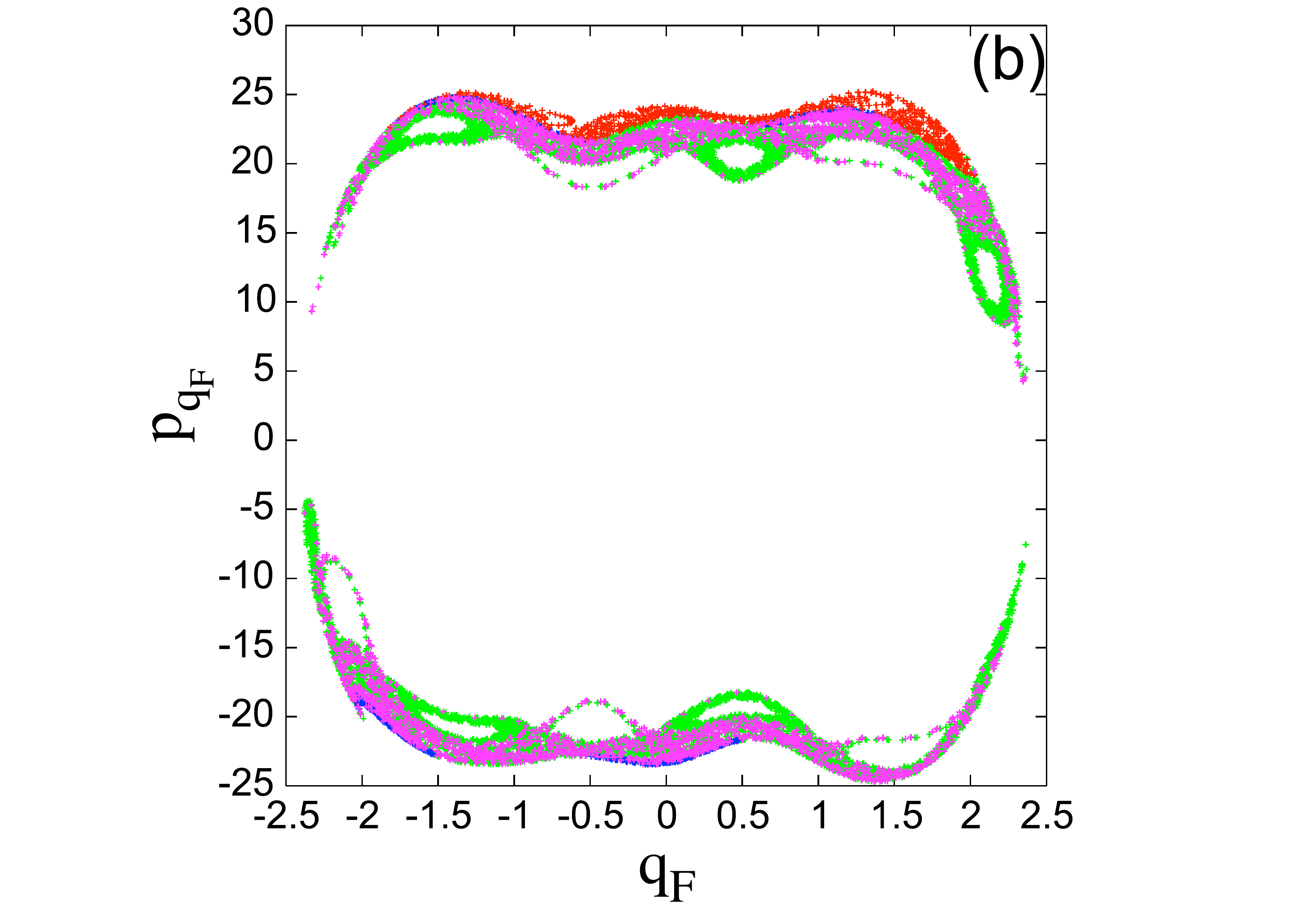}
\includegraphics[scale=0.3]{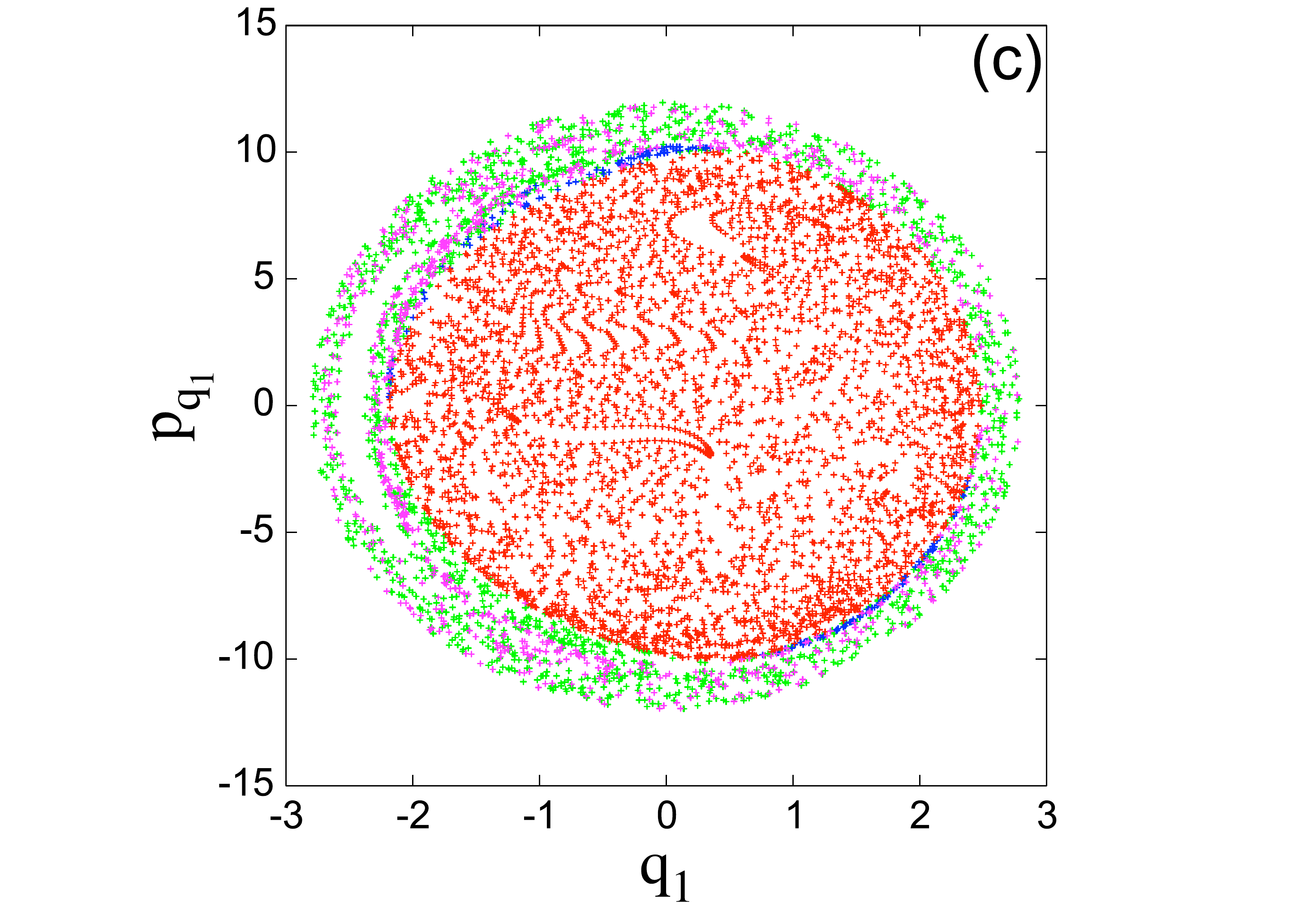}
\includegraphics[scale=0.3]{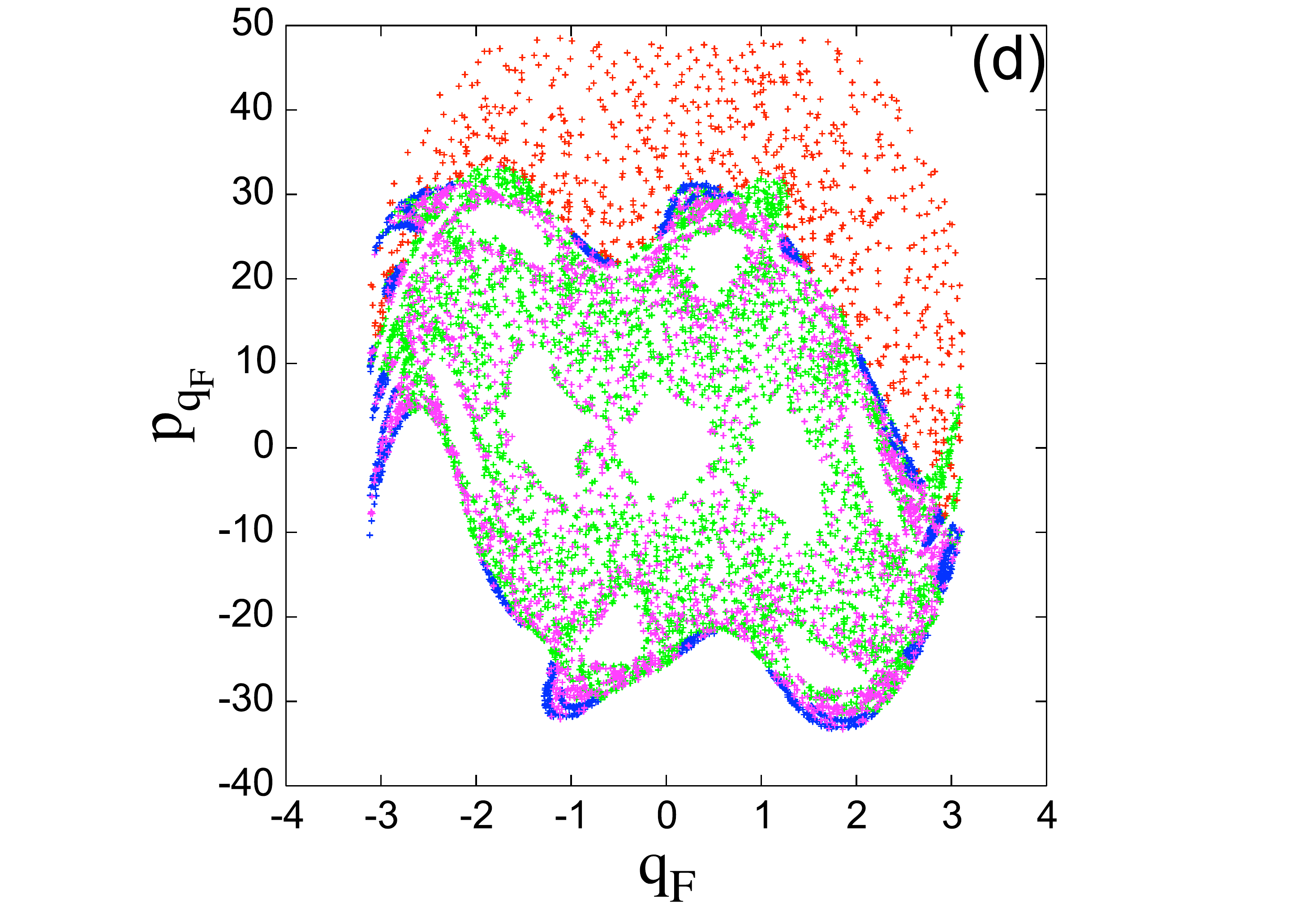}
\caption{Poincar\'e surfaces of sections (PSSs). (a) PSS$_1$ at $E=0.010$ $E_h$. (b) PSS$_F$ at $E=0.010$ $E_h$. (c) PSS$_1$ at $E=0.040$ $E_h$.
(d) PSS$_F$ at $E=0.040$ $E_h$.}
\label{fig:8} 
\end{figure}

We have used two different PSS which give
complementary views of the dynamics. The first one (PSS$_1$) was defined by fixing $q_F=0$ and 
recording points when they cross this plane with a momentum $p_{q_F}>0$. The second one (PSS$_F$)
was chosen at $q_1=0$ with momentum $p_{q_1}>0$. These two PSS are depicted in Fig~\ref{fig:8} 
for two different energies. Panel (a) is PSS$_1$ for energy $E=0.01$ Hartree above oxirene minimum and panel (b) is PSS$_F$ for
the same energy. Panels (c) and (d) are the equivalent PSS at energy $E=0.04$ Hartree above oxirene minimum.

For each panel we have used different colours for points on the PSSs belonging to trajectories of a different category as were classified above.
The colours are consistent with those in the previous plots: 
hence, red dots denote points belonging to direct reactive trajectories, green dots
denote the roaming reactive ones, blue the direct non-reactive and magenta 
the roaming non-reactive trajectories, respectively. 
In  Fig~\ref{fig:8} we see a clear separation between the red and the other colours with the red concentrated in the middle
of the figure. This is consistent with the fact that 
direct reactive trajectories have their energy
mainly distributed on the ($q_F,p_{q_F}$) DoF. Interestingly, the blue non-reactive direct trajectories 
appear at the boundary between the red and the roaming trajectories.
Such reactivity  boundaries have been observed in many studies and in several different
contexts (see for example \cite{MauguiereRoaming,Nagahata13,Grice87,Mauguiere13}, an references therein). 

On the other hand, the green and magenta points (roaming trajectories)
cover the periphery in the plots of PSS$_1$ indicating that more energy 
is involved in the ($q_1,p_{q_1}$) DoF for these types of
trajectories. The two plots of PSS$_F$ for the two energies show that the green and magenta dots 
are concentrated around some islands 
of stability (blank area). These blank areas correspond to the stable
regions in phase space associated with the stable resonant POs. 
The large blank area in panel (b) and for the energy $E=0.01$ Hartree is not accessible to trajectories initiated on the DS at EP$_1$. 
However, as energy increases to 0.04 Hartree this stable area shrinks and the three central blank domains 
in panel (d) are attributed to trajectories in the vicinity 
of the three minima, which correspond to oxirene (EP$_4$) 
and the two formylmethylene minima (EP$_2$, EP$_6$). 

\subsubsection{Classical autocorrelation functions}
\label{CAF}

Quantum calculations of the rate constant of ketene isomerization have shown 
that there are resonant features that are spaced at 70-80 cm$^{-1}$ apart 
and with a width of about 10 cm$^{-1}$. 
These resonances have been attributed to Feshbach energy-transfer, 
or dynamical resonances
that occur at energies above the barrier to isomerization \cite{gezelter1995resonant}. 
According to Gezelter and Miller \cite{gezelter1995resonant}
``these dynamical resonances appear because the multimode potential energy surface has a strongly bent region 
between the two outer transition states, which acts as a dynamical bottleneck even at total energies above the outer transition state energies''.
The present classical trajectory results 
suggest that these resonances can be identified with the resonant periodic orbits, 
which are the result of a cascade 
of center-saddle bifurcations shown in Fig. \ref{fig:7} \cite{Farantos09}. 

This notion can be verified by calculating the classical analogue ($\Omega(t)$) of 
the quantum survival probability function ($|C(t)|^2$) for a wavepacket centered on the classical PO.
In this case, the correspondence between spectrum and phase space structure is relatively straightforward. 
We can pass from the quantum to the classical analogue of  the autocorrelation function by replacing the trace in the density matrix
given in Eq. \ref{ACFQ} with an integral over the phase space, and by replacing the density operators  
($|\phi(q,t)\rangle\langle\phi(q,t)|$) with classical distribution functions, 
$\rho(q,p)$, \cite{Baranger1958,Heller1980}.
\begin{equation}
\label{ACFC}
\Omega(t)=\int\rho[q(0),p(0)] \rho[q(t),p(t)]dqdp. 
\end{equation}
\begin{equation}
\label{ACFQ}
|C(t)|^2=|\langle \phi(q,0)|\phi(q,t)\rangle|^2 = 
\langle \phi(q,0)|\phi(q,t)\rangle \langle \phi(q,t)|\phi(q,0)\rangle ,
\end{equation}

In our case the classical initial distribution $\rho[q(0),p(0)]$  is a Gaussian function of coordinates and momenta. 
The spectrum is then defined as the Fourier transform, 
\begin{equation}
I_c(\omega)=\int e^{i\omega t} \Omega (t)dt.
\end{equation}
The classical survival probability function, Eq. \ref{ACFC}, and its Fourier transform  
have been used in studies of molecular spectroscopy and dynamics  
by Gomez Llorente, Pollak, and Taylor \cite{Pollak1989,Taylor1989}.

\begin{figure}
\includegraphics[scale=0.5]{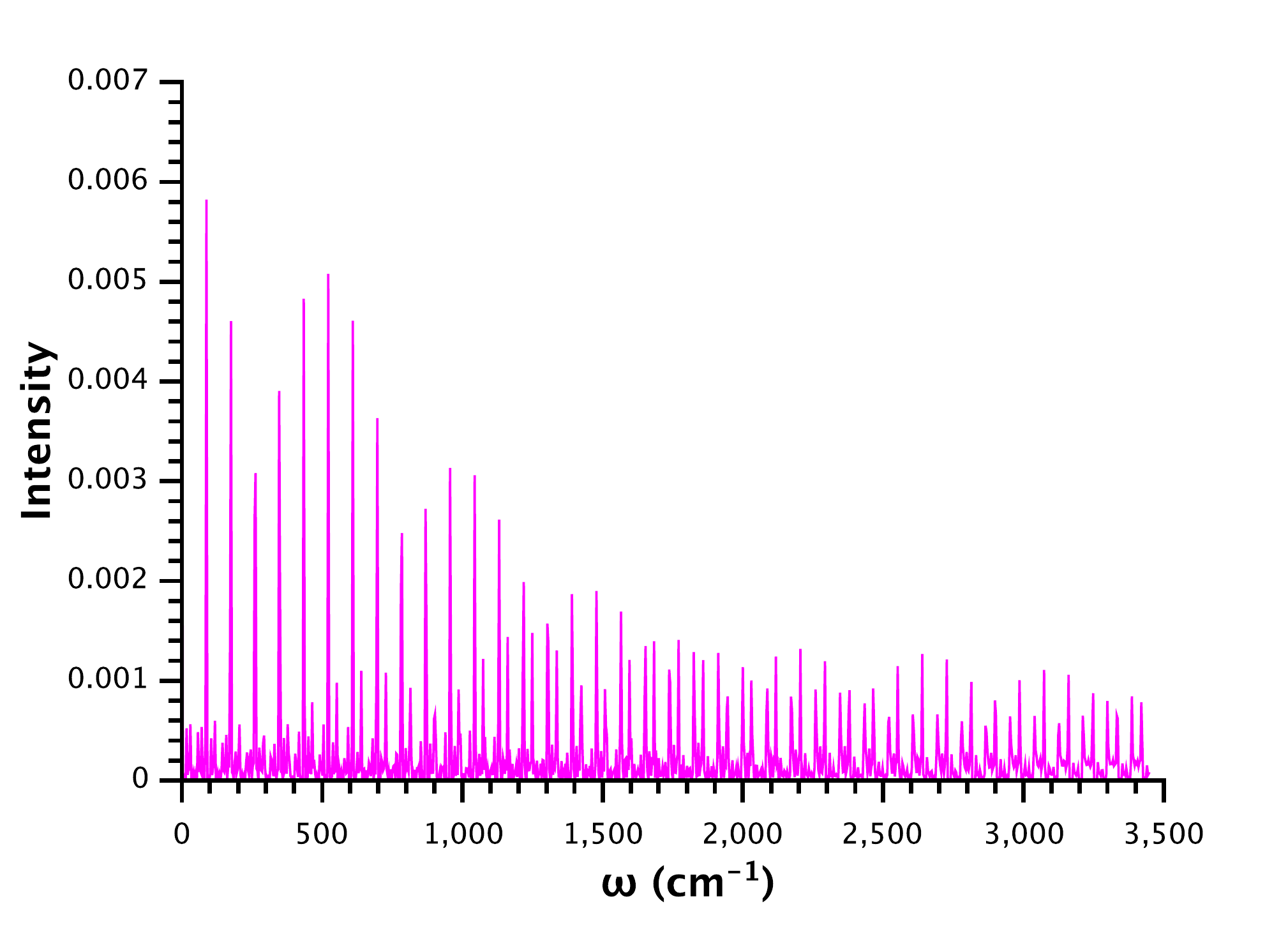}
\caption{Autocorrelation function for trajectories in the neighbourhood of 1:6 periodic orbit at energy of 0.04 $E_h$.}
\label{fig:9} 
\end{figure}

Figure \ref{fig:9} shows a typical spectrum produced by a batch of 
1000 trajectories selected from a Gaussian distribution
centered on the stable 1:6 PO and at energy 0.04 $E_h$. 
The spacing of eigenfrequencies is 83 cm$^{-1}$  and the regularity of 
the spectrum reflects the regularity of the region of phase space in which the trajectories are trapped. 
Higher order POs, like the 1:7, have longer periods and appear at energies closer to the isomerization saddle. 
Their frequencies match the quantum mechanical frequencies in the interval of 70-80 cm$^{-1}$ found by Gezelter and Miller
\cite{gezelter1995resonant}. 
 
\section{Phase space reaction pathways and roaming}
\label{sec:4}

Reaction pathways are conventionally 
identified in nuclear configuration space by minimum energy paths on the 
potential energy surface, since the latter is the fundamental concept 
used to interpret chemical reactivity. 
Thus, the presence of multiple saddles on the PES suggests the existence of multiple reaction pathways. 
However, the development of non-linear mechanics has enabled the rigorous definition of 
objects like reaction pathways and transition states in phase space for polyatomic molecules.

The strategy followed to understand the motions of non-linear dynamical systems is to reveal the geometry of phase space in a systematic way
by first locating equilibria, and from them the emanating principal families of periodic orbits, 
tori, NHIMs and (un)stable manifolds associated with NHIMs. 
The latter are key protagonists in controlling reaction fluxes. 
Both for the isomerization dynamics of ketene as well as in our previous 
work on the Chesnavich model for ion-molecule reactions 
we have shown that, by defining transition states using the 
appropriate NHIMs (unstable periodic orbits for 2D molecular models),  
ordinary (MEP) and roaming reaction pathways emerge within a phase space framework. 
Similarities between the two type of reactions
emphasized in subsection \ref{sec:3.2}; we now point out some significant
differences.

Ketene belongs to the general type of molecules for which NHIMs are associated with potential energy saddles. 
The Chesnavich model for ion-molecule reactions is a model 
which manifests the existence of NHIMs not associated with potential
energy saddles. Indeed, as  shown in refs \cite{Mauguiere2014282,MauguiereRoaming}, 
the loose transition state as well as the
tight one are periodic orbits originated from center-saddle bifurcations. 
The loose transition state is a rotating type periodic orbit (relative equilibrium). 
The tight transition state is a periodic orbit associated with the 
barrierless MEP for association/dissocation. Hence, for these type of
reactions the phase space is mandatory to define reaction pathways, 
since the potential energy alone is inadequate.  

By attributing roaming to trapped trajectories in the region of non-linear resonances in phase space, 
we are able to generalize the 
concept of roaming. From the PSS in Fig. \ref{fig:8} and in the low energy regime (0.01 Hartree), 
the roaming pathways are determined by the resonances 1:6 and 1:7 as indicated by the types of POs in Fig. \ref{fig:7}. 
From the large number of trajectories run none penetrates close to the regions of formylmethylene and oxirene. 
At higher energies (0.04 Hartree) the regular phase space region 
governed by the equilibria, EP2, EP4 and EP6, shrinks and the reactive trajectories initiated 
from the NHIMs of the
outer saddle penetrate into this region. 
In this way, we infer that the new resonances associated with these minima define new roaming 
reaction  pathways. It is worth mentioning that non-linear resonances associated with center-saddle bifurcations of periodic orbits have
been found to be a general phenomenon in molecular dynamics \cite{Farantos09}.

Finally, it is essential to stress that, although roaming as an 
alternative reaction pathway was identified  in the dissociation of formaldehyde 
with long range interactions between the fragment species, 
several other molecules with competing non-MEP have previously been studied 
\cite{Hase1989,Hase2002,Hase2007,Hase2008,Hase2010,Carpenter04,Lourderaj09}. 

\section{Summary and conclusions}
\label{sec:5}

In this paper we have used a reduced dimensional model initially proposed by Gezelter and Miller to study the isomerization
dynamics of Ketene molecule. The study of Ulusoy \textit{et al.} \cite{ulusoy2013effects,Ulusoy13b}
discussed the connection of ketene isomerization
and roaming reactions mechanism.  The question of how our previous interpretation of roaming phenomenon
\cite{Mauguiere2014282,MauguiereRoaming} fits the ketene isomerization situation naturally arose.

It was found that the previous interpretation of the
roaming mechanism in terms of a trapping mechanism of trajectories between
two DSs enhanced by non-linear resonances also fits the ketene isomerization reaction well. We were
able to classify classical trajectories into qualitatively different types of trajectories 
and compute fractions of different types of trajectories; this data could 
be used to correct the flux in a reaction rate calculation. 
As in our previous study, the question of statistical
dynamics was raised and investigated by gap time analysis. 
Significant deviation from the statistical assumption of TST was found. 
In addition, evidence of the trapping mechanism of the trajectories 
by resonant POs has been given using surfaces of section. 

The importance of ketene isomerization stems from the experimental investigations of
Moore and co-workers. Furthermore,
the quantum mechanical results of Gezelter and Miller are in overall 
agreement to the experimental results.
Our phase space analysis provides further insight into this important system; 
for example, ketene provides an experimental
manifestation of the importance of the center-saddle bifurcations of periodic orbits. 

In the present calculation and for the energies studied the direct type reactive trajectories dominate (Fig. \ref{fig:6}).
However, if oxirene could be isolated and photoexcited into resonant states, 
the roaming branch of the reaction would then be the
dominant one. 
In other words, knowledge of the phase space structure in a particular range of 
energies in principle enables us to extert some degree of control over chemical
reaction pathways. Analysis of classical phase space as presented here can achieve 
a degree of resolution not necessarily accessible 
to experiment. Nevertheless, experimental techniques 
are continually advancing and methods may be developed in the future to
control roaming pathways \cite{Ashfold06}.

\begin{acknowledgements}
We wish to thank Dr.\ D. Gezelter, Prof.\  W. H. Miller and Prof.\ R. Hernandez and coworkers 
for helpful discussions.
This work is supported by the National Science Foundation under Grant No.\ CHE-1223754 (to GSE).
FM, PC, and SW  acknowledge the support of the  Office of Naval Research (Grant No.~N00014-01-1-0769),
the Leverhulme Trust, and the Engineering and Physical Sciences Research Council (Grant No.~ EP/K000489/1).
\end{acknowledgements}

\bibliographystyle{spphys}       

%
%

\newpage

\newpage

\newpage

\newpage

\newpage

\newpage

\begin{figure}
\includegraphics[scale=1]{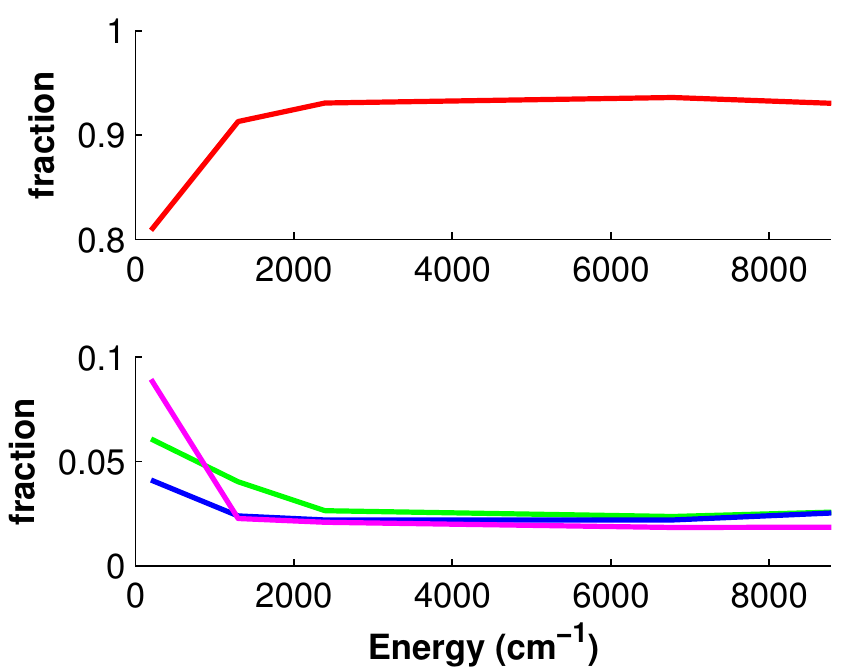}
\caption{Fractions of the different trajectories categories as a function of the energy above the barrier for isomerization. The different fractions are computed with the same number of trajectories as indicated in Fig~\ref{fig:4}.}
\label{fig:6} 
\end{figure}

\newpage

\begin{figure}
\includegraphics[scale=1]{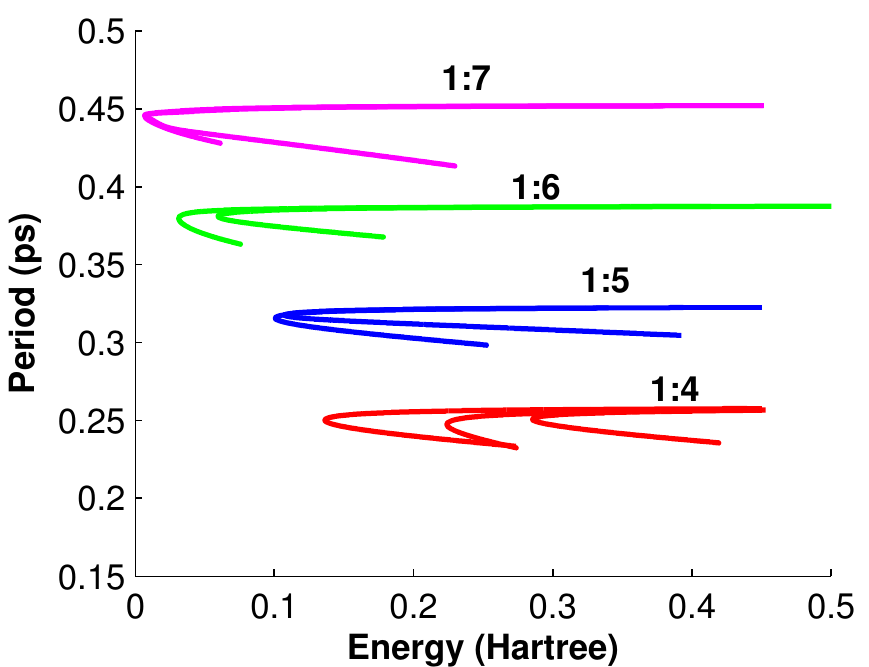}
\caption{Continuation/Bifurcation diagram for the resonant periodic orbits in the roaming region.}
\label{fig:7} 
\end{figure}

\newpage

\newpage

\end{document}